\documentclass[10pt,onecolumn,titlepage]{article}
\usepackage{scicite}
\usepackage{times}
\usepackage[utf8]{inputenc}
\usepackage{graphicx}
\usepackage{amsmath}
\usepackage{amssymb}
\usepackage{color}

\newenvironment{sciabstract}{%
\begin{quote} \bf}
{\end{quote}}

\newcounter{lastnote}

\title{Response to Comment on ``The role of electron-electron interactions in two-dimensional Dirac fermions"}

\author
{Ho-Kin Tang,$^{1,2}$ J. N. Leaw,$^{1,2}$ J. N. B. Rodrigues,$^{1,2}$ I. F. Herbut,$^{3}$\\ P. Sengupta,$^{1,4}$ F. F. Assaad,$^{5}$ and S. Adam$^{1,2,6,\ast}$\\
\\
\normalsize{$^{1}$Centre for Advanced 2D Materials, National University of Singapore,}\\
\normalsize{6 Science Drive 2, 117546 Singapore.}\\
\normalsize{$^{2}$Department of Physics, Faculty of Science, National University of Singapore,}\\
\normalsize{2 Science Drive 3, 117542  Singapore.}\\
\normalsize{$^{3}$Department of Physics, Simon Fraser University,}\\
\normalsize{Burnaby, British Columbia V5A 1S6, Canada.}\\
\normalsize{$^{4}$School of Physical and Mathematical Sciences, Nanyang Technological University,}\\
\normalsize{21 Nanyang Link, 637371 Singapore.}\\
\normalsize{$^{5}$Institut f\"ur Theoretische Physik und Astrophysik, Universit\"at W\"urzburg,}\\
\normalsize{Am Hubland, D-97074 W\"urzburg, Germany.}\\
\normalsize{$^{6}$Yale-NUS College, 16 College Ave West, 138527 Singapore.}\\
\\
\normalsize{$^\ast$To whom correspondence should be addressed; E-mail: shaffique.adam@yale-nus.edu.sg}
}

\date{}

\begin{document}
\maketitle
\begin{sciabstract}
Hesselmann {\it et al}.~question one of our conclusions, namely, the suppression of Fermi velocity at the Gross-Neveu critical point for the specific case of vanishing long-range interactions and at zero energy.  The possibility they raise could occur in any finite-size extrapolation of numerical data.  While we cannot definitively rule out this possibility, we provide mathematical bounds on its likelihood.
\end{sciabstract}

\noindent Hesselmann {\it et al}.~question the procedure we use to extract the interaction correction to the Fermi velocity $\Delta v_F(k)$ from the quantum Monte Carlo (QMC) numerical simulations.  To be clear:  They discuss the case of no long-range Coulomb interactions ($\gamma = 0$), and very close to criticality ($U \rightarrow U_c$).  To put this in context, our Fig.~2 in Tang {\it et al}.~\cite{Tang2018} comprises about 120 data points for on-site interactions ranging from weak to strong coupling, and with varying long-range interaction.  Their critique concerns at most 3 of these data points.  Additionally, our data can be thought of as momentum slices of Fig.~2, for which we use 16 such slices.  Hesselmann {\it et al}.~restrict their criticism only close to the Dirac point ($\Delta k \rightarrow 0$), and as such, concern only these three data points out of the $\sim$2000 projected data sets, and therefore no more than 0.2 percent of the QMC data in Ref.~\cite{Tang2018}.  Most of our core findings and conclusions are unaffected by the concerns they raise. For the rest of this reply, we restrict ourselves to the $\gamma =0$, $U \rightarrow U_c$, and $\Delta k \rightarrow 0$ limit that interests Hesselmann {\it et al}. We consistently use the following estimator for $\Delta v_F$: 
\begin{equation}
    \Delta v_F^{\rm {Tang}}(k) = \lim_{L \rightarrow \infty} \frac{\Delta E \left(k, L \right) - \Delta E \left( 0, L \right)}{k}, \\
\end{equation}
where $\Delta E \left(k, L \right)$ is obtained from the QMC data for system size $L$.  We evaluate our estimator at $k_{\rm min}= 4 \pi/(\sqrt{3} L_{\rm max})$ which is the smallest momentum accessible in the Monte Carlo method.  $\Delta E \left(k_{\rm min}, L \right)$ is obtained for $L < L_{\rm max}$ from $\Delta E(k_L, L)$ and $\Delta E(0, L)$ by linear interpolation (see Fig.~1). Here $k_{\rm L} =4 \pi/(\sqrt{3} L)$.  Notice that if we could simulate lattices of infinite size, then our estimator would be identical to the mathematical definition of $\Delta v_{F} (k \rightarrow 0)$   
\begin{equation}
    \Delta v_F^{\rm {Tang}}(k \rightarrow 0) = \lim_{k_{\rm min} \rightarrow 0} \lim_{L \rightarrow \infty} \frac{\Delta E \left(k_{\rm min}, L \right) - \Delta E \left( 0, L \right)}{k_{\rm min}}.
\end{equation}
\noindent Hesselmann {\it et al}.~observe from our numerical data that close to criticality, the Dirac point energy is more strongly affected by finite lattice size than neighboring momenta, and attribute this to the persistence of the antiferromagnetic order parameter in the semi-metallic regime.  To correct for this, they outline an alternate procedure: (i) set the Dirac point energy to zero (throwing away all information that the QMC provides about the Dirac point), and (ii) use an estimator obtained from a single system size
\begin{equation}
    \Delta v_F^{H} = \frac{\Delta E \left( \frac{4 \pi}{\sqrt{3} L_{\rm max}}, L_{\rm max} \right)}{4 \pi/(\sqrt{3} L_{\rm max})},
\end{equation}
\noindent that ignores any finite-size scaling information available within the QMC data.  Figure~1 shows that the two estimators give different results when applied to our QMC data. The purpose of this note is to explain why. \\  

\noindent To begin, notice that in the thermodynamic limit, the Hesselmann {\it et al}.~estimator is different from the mathematical definition of the Fermi velocity
\begin{equation}
     \Delta v_F^{H} = \lim_{k_{\rm min} \rightarrow 0} \frac{\Delta E \left( k_{\rm min},  \frac{4 \pi}{\sqrt{3} k_{\rm min}} \right)}{k_{\rm min}} \neq \Delta v_F^{\rm {Tang}}(k\rightarrow 0).
\end{equation}
\noindent  This is illustrated graphically in the inset to Fig.~1.  $\Delta v_F^{H}$ is taken along the black diagonal arrow, while $\Delta v_F^{\rm {Tang}}(k\rightarrow 0)$ is taken along the red horizontal arrow.  We note that if in the thermodynamic limit the two estimators disagree, then ours is always correct.  However, at issue here is not the thermodynamic limit, but the finite lattice sizes achievable using quantum Monte Carlo.  Although we can not make  {\it a priori} assumptions about the functional form of $\Delta E(k, L)$ at the critical point (since we have a strongly correlated many-body state), we can still construct hypothetical functions $\Delta E_j(k, L)$ to illustrate when and why $\Delta v_F^{\rm {Tang}}(k_{\rm min})$ and $\Delta v_F^{H}$ disagree. \\

\noindent (1) First consider $\Delta E_1(k, L) = \Delta v_F^{\rm True}(k)k + \alpha(k)/L$, where $\Delta v_F^{\rm True}(k) = \Delta v_0 + \Delta v_1 k$, (with $\Delta v_0 = -0.3$ and $\Delta v_1 = 0.1$ chosen to be consistent with the QMC data), and $\alpha(k) = \alpha_0 + \alpha_1 k$, (where $\alpha_0 = 2$ and $\alpha_1 = 1$ are similarly consistent with the data). For $L=24$, this gives $\Delta v_F^H \approx 0$, and $\Delta v_F^{\rm Tang}(k_{\rm min}) \approx  \Delta v_F^{\rm True} = -30\%$.  This simple and reasonable construction shows how it is possible for the Hesselmann {\it et al}.~estimator to find no Fermi velocity renormalization, despite there being a strong suppression correctly captured by our estimator (see also the inset Fig.~1B). \\

\noindent (2) Now consider $\Delta E_2(k, L) = \alpha_0 \delta(k)/L + \Delta v_F^{\rm True} k$, where $\delta(k)$ is the Dirac delta function.  This is an extreme example of Hesselmann {\it et al.'s}~concern: only the Dirac point has finite-size effects, but no other momenta.  We must emphasize that this functional form is inconsistent with our numerical data.  Nonetheless, for this hypothetical worst case scenario, $\Delta v_F^H =  \Delta v_F^{\rm True}$, and $\Delta v_F^{\rm{Tang}}(k_{\rm{min}}) = \Delta v_F^{\rm True}-\sqrt{3}\alpha_0/(4\pi)$.  Taking $\Delta v_F^{\rm True} = 0$, and $\alpha_0$ as above, we would underestimate $v_F$ by at most 28\%. \\

\noindent Hesselmann {\it et al}.'s core claim is that ``the strong suppression of Fermi velocity ... near the Gross-Neveu quantum critical point {\it merely} reflects the enhanced finite-size effects ... at the Dirac point, but {\it not} the renormalization of the actual low-energy dispersion".  Since Hesselmann {\it et al}.~cannot exclude $\Delta E_1(k, L)$ as a possible energy function, {\it their claim is unsubstantiated}.  Moreover, even in the hypothetical worst-case scenario $\Delta E_2(k, L)$, for the data in Fig.~1, it would require that $\alpha_0 > 2.79$ for their claim to be correct.  As seen in Fig.~1C, our QMC data lies outside the shaded region, and therefore, for this case, their {\it claim is false}.  In addition, the finite-size scaling at non-zero momenta (e.g. Fig.~1D) and the observation that all the data points for $L<24$ lie {\it below} $\Delta v_F^{H}$ provides {\it clear and convincing evidence} that a functional form like $E_1(k,L)$ is more likely than $E_2(k,L)$. \\

\noindent Some further remarks are in order: \\

\noindent (a) The positive $\Delta v_F^H$ is counter intuitive from a physical point of view as the fermions will scatter off paramagnons and thereby slow down.  This interaction with a bosonic mode is analogous to graphene interacting with phonons for which $v_F(k)$ is suppressed close to the Dirac point and enhanced for energies larger than the Debye energy (e.g. Fig.~2 of Ref.\cite{Bostwick2006}). This framework allows us to understand how a functional form like $\Delta E_1(k,L)$ arises physically, and why $\Delta v_{F}^H$ incorrectly gets an enhanced Fermi velocity.  \\

\noindent (b) The renormalization group flows in Ref.~\cite{Tang2018} were most strongly influenced by the logarithmic divergence (at finite $\gamma$)  of $\Delta v_F(k)$ at {\it large} momenta, and as such, the numerical value of $\Delta v_F$ at $\gamma = 0$, $U=U_c$ and $k \rightarrow 0$ is not germane to our paper.    Actually we did not even claim in the paper to be the first to calculate it.  In Fig.~14 of Ref.~\cite{Otsuka2016}, they show a 38\% suppression of $\alpha(v_{\rm F})$, which is the pre-factor of the density-density correlation function (in the Brinkmann-Rice metal-insulator transition, both $v_F$ and $\alpha$ vanish at the transition).  Moreover, the Fermi velocity renormalization can be obtained from the specific heat ($c_v \sim T^2 / v_F^2$), for which Ref.~\cite{Paiva2005} in Fig.~13 calculates a $\sim30\%$ enhancement of $c_v$ at $U=U_c$.  Both these works (and ours) suggest a velocity renormalization by using a finite-size scaling on a honeycomb lattice with local hopping.  By contrast, for a model where the non-interacting fermions have long-range hopping, Ref.~\cite{Thomas2018} uses Eq.~3 (without finite-size scaling) and find no renormalization. It remains undetermined as to whether this discrepancy is intrinsic or due to a different choice of estimator.  \\

\noindent (c)  Another indication that $E_1(k,L)$ is more likely than $E_2(k,L)$ is the relative stability of the two estimators.  We could use $\Delta v_F^H$ for our QMC data, and our main conclusions would not change except at $U \rightarrow U_c$ (away from criticality, $\Delta v_F^{\rm Tang} = \Delta v_F^H$ in the thermodynamic limit).  However, this presents several problems: (i) $\Delta v_{F}^H$ is inconsistent with the actual QMC data; (ii) We would need different fitting procedures for different parts of our phase diagram; and (iii) $\Delta v_{F}^H$ is unstable with changing $L$.  Going from $L=15$ (data in our paper) to $L=24$ (data available since publication), $\Delta v_F^{H}$ changes from $-1.17\%$ to $+2.94\%$, while $\Delta v_F^{\rm{Tang}}$ only changes from $-31.4\%$ to $-30.5\%$ (see Fig. 1). \\

\begin{figure}
    \centering
    \includegraphics[width=1\linewidth]{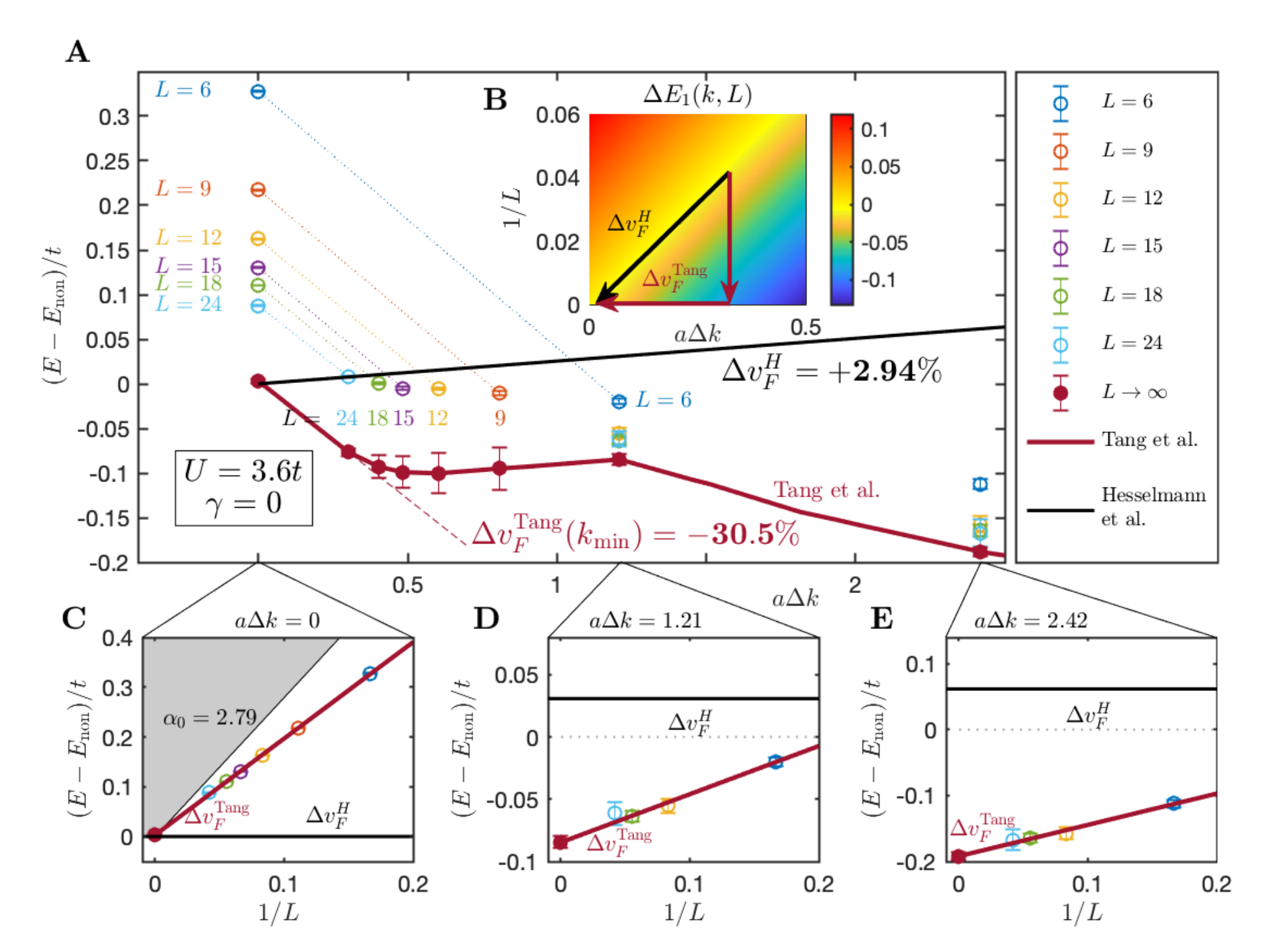}
    \caption{Main panel shows the change in energy for Dirac fermions due to electron-electron interactions as determined by the projective quantum Monte Carlo method developed in Ref.~\cite{Tang2018}.  Open data points are for lattice sizes $L\times L$, where $L=6,9,12,15,18,24$.  The solid red line is our estimator $\Delta v_F^{\rm Tang}(k)$ (Eq.~1), while the black line is the alternate estimator $\Delta v_F^H$ (Eq.~3).  The two estimators disagree at $k_{\rm min}$: $\Delta v_F^{\rm Tang}(k_{\rm min}) = -30.5\%$ while $\Delta v_F^H = +2.94\%$. Inset B (explained in the text) shows that our estimator is always correct in the thermodynamic limit. Outset C shows the finite-size scaling of the Dirac point.  Since the QMC data is outside the shaded region, even in the hypothetical worst-case for our estimator, our numerical data is inconsistent with an unrenormalized Fermi velocity.  Outsets D and E show finite-size scaling of $v_{F} (k)$ at non-zero momenta providing clear evidence that $E_1(k,L)$ (the best case for our estimator) is more likely than $E_2(k,L)$ (the worst case for our estimator). }
    \label{fig:my_label}
\end{figure}

\newpage 

\bibliographystyle{Science}
\bibliography{biblio}

\end{document}